\documentclass[aps,prl,twocolumn,groupedaddress]{revtex4}

\usepackage{amsmath}
\usepackage{hyperref}

\newcommand{\noperp}{{\phantom \perp\!}}

\begin{document}


\title{Acausality of Massive Gravity}


\author{S.~Deser}
\email[]{deser@brandeis.edu}
\affiliation{Lauritsen Lab, Caltech, Pasadena CA 91125 and Physics Department, Brandeis University, Waltham, MA 02454, USA}

\author{A.~Waldron}
\email[]{wally@math.ucdavis.edu}
\affiliation{Department of Mathematics, University of California, Davis, CA 95616, USA}


\date{\today}

\begin{abstract}
We show, by analyzing its characteristics, that the ghost-free, 5 degree of freedom,  Wess--Zumino  massive gravity model 
admits superluminal shock wave solutions and thus is acausal.
Ironically, this pathology arises from the very constraint that removes the (sixth) Boulware-Deser ghost
mode.
\end{abstract}


\maketitle

\section{Introduction}

Over four decades ago, Isham, Salam and Strathdee proposed a 2-tensor ``$f$-$g$'' theory~\cite{Salam}
by adding to the Einstein action that of a second vierbein~$f_{\mu}{}^m$,  plus
a nonderivative coupling term, leaving  a single common coordinate
invariance. Of particular interest is the limit of nondynamical (say flat)~$f$,
giving gravitons a finite range due to the coupling ``mass'' term. It
was rapidly shown~\cite{BD} however, that unlike their linearized massive spin~2
Fierz-Pauli~(FP) limits, these models suffered from a ghost problem: generic
nonlinearities reinstate a 6th degree of freedom~(DoF), beyond the linearized
$2s+1=5$ DoF, one of which is necessarily ghostlike. A final twist, also from that
time, was the Wess--Zumino~\cite{WZ} discovery of a distinguished set of~$f$-$g$
mass terms of which at least one is  immune from this disease, keeping~5~DoF.  Because~\cite{WZ} was
only published without detail in lecture notes, it remained unknown. Separately, other analyses showed that the linearized
theory's matter coupling seemed to suffer a ``vDVZ'' zero-mass discontinuity~\cite{vDVZ},
as well as a failure of the Birkhoff theorem~\cite{PvN}. Hence,
the subject remained moribund until the recent (independent) rediscovery~\cite{deRham}
of the results~\cite{WZ} plus two new $f$-$g$ models. This exhumation has, unsurprisingly, generated an immense
industry (see the recent survey~\cite{HB}). Our purpose is to re\-inter~{$f$-$g$}. 
We will show that the 5 DoF, Wess-Zumino model is acausal~\footnote{
An earlier massive gravity causality study~\cite{Gruzinov} found superluminal behavior in the auxiliary fields of the model's St\"uckelberg formulation.
However, these superluminal modes amount to unphysical background metrics~\cite{deRhamS}. Nonetheless, this effect could well be related to known  horizon
and null energy difficulties of one of either one of the metrics  of a bimetric theory~\cite{Jacobson}. We emphasize that our classical results are valid for all (nonzero)~$m$; hence, even though these---like any classical---models have a limited  ranged of validity (as has been argued from a quantum viewpoint in~\cite{Kaloper}), their acausality is surely present in that region.}.
Our methods also show that
of the two remaining~5~DoF models~\cite{deRham}, one is definitely acausal and the other likely so~\cite{DSW}. Paradoxically, acausalities arise precisely because of the very constraint that
removes the ghost. Note that there is no conflict  between acausality
and ghostlessness,
as witnessed by  the old ``charged'' higher ($s>1$) spin  interactions with
Maxwell and gravity,
say those of $s=(3/2,2)$~\cite{Velo,Shamaly,DW}, that are also invalidated only by acausality.

Our results will be obtained by using the method of characteristics, analyzing
the constraints' shock wave discontinuities, in particular, that of the ``fifth'' scalar one that
results from combining the trace and double divergence of the field equations, just as is done in the
linear FP model, to find a derivative-free constraint. 

\section{The Model and the Fifth Constraint}

Our concrete 5 DoF model is
\begin{equation}\label{EOM}
{\cal G}_{\mu\nu}:=G_{\mu\nu}(g)+
m^2 \big(f_{\mu\nu}-g_{\mu\nu} f\big)=0\, ,
\end{equation}
where all indices are moved by the dynamical metric $g_{\mu\nu}$ and its associated
vierbein $e_\mu{}^m$;
in particular $f_{\mu\nu}$ is the fixed background vierbein $f_{\mu}{}^m$ times $e_{\nu m}$,
and is manifestly symmetric on-shell. Vanishing of its antisymmetric part yields six conditions. 
Taking the reference $f_\mu{}^m$  field as the flat bein is a popular choice but is not physically required; in fact,  our results, both for acausality and the absence of the sixth ghost mode, depend neither on $f$ being flat nor  the dimensionality of spacetime.
The parameter $m^2$ reduces to
the FP mass in the weak $e$-field limit.
Next, we proceed as in the FP development and seek five constraints to
reduce the {\it a priori} ten metric
DoF (now that coordinate invariance is lost due to the preferred
background). The single derivative, four-vector, constraint
is obviously (by the Bianchi identity) the covariant $g$-divergence of Eq.~(\ref{EOM}),
$$
0={\cal C}_\nu:=\nabla^\mu {\cal G}_{\mu\nu}=m^2 \big(\nabla.f_\nu - \nabla_\nu f\big)\, .
$$
The scalar
constraint results from taking the (covariantized)
FP combination
\begin{equation}\label{V}
0={\cal C}:=\nabla_\mu\big(\ell^{\mu\nu}\nabla.{\cal G}_\nu\big)+\frac{m^2}2\,  {\cal G}
\end{equation}
with $\ell^{\mu\nu}:=\ell^\mu{}_m e^{\nu m}$, where $\ell^\mu{}_m$ is the inverse of the background vierbein~$f_\mu{}^m$.
The proof that~${\cal C}$ is indeed a constraint, {\it i.e.}, devoid of second
derivatives, is simple: following~\cite{Deffayet},
we observe that the (torsion-free) Levi--Civita spin connection $\omega(e)_\mu{}^m{}_n$ corresponding to the vierbeine~$e_\mu{}^m$
will in general become torsionful if employed as the spin connection for the nondynamical vierbeine $f_\mu{}^m$. The difference between this connection
and the Levi--Civita spin connection $\omega(f)_\mu{}^m{}_n$ of $f_\mu{}^m$ yields the contorsion tensor 
$$
K_\mu{}^m{}_n:=\omega(e)_\mu{}^m{}_n-\omega(f)_\mu{}^m{}_n\, .
$$
It measures the failure of parallelograms of the dynamical metric to close with respect to the background metric  (and {\it vice versa}).
As will become apparent, it is important to emphasize that flatness of the background metric does {\it not} ensure vanishing  
contorsion. In these terms, the vector constraint reads $$0={\cal C}_\mu = m^2 K_\nu{}^{\nu\rho}f_{\mu\rho}\, .$$
In particular, this means that metric derivatives enter the vector constraint only through the trace of the spin connection $\omega(e)$. However the
leading (second) derivative terms of the scalar curvature~$R$ are proportional to $\partial_\mu \omega(e)_\nu{}^{\nu\mu}$. Hence the linear combination of the divergence of the vector constraint and the trace of the equation of motion quoted in Eq.~(\ref{V}) yields the remaining scalar constraint~${\cal C}=0$. This ensures that the model does not propagate spurious
ghost degrees of freedom and thus evades the generic difficulties associated with massive gravity theories~\cite{BD}. 

For our purposes an explicit evaluation of the scalar constraint~${\cal C}$ is needed: we first express the scalar curvature  in terms of the contorsion
\begin{equation*}
\begin{split}
R\ &=\ 2\, \nabla_\mu K_\nu{}^{\nu\mu}-K_{\mu\nu\rho}K^{\nu\rho\mu}-K_\mu{}^{\mu\rho}K_\nu{}^{\nu}{}_{\rho} \\[1mm]&+\ e^\nu{}_me^\mu{}_n R(f)_{\mu\nu}{}^{mn}\, ,
\end{split}
\end{equation*}
where $R(f)$ is the Riemann tensor corresponding to the vierbeine $f_\mu{}^m$. Observing that the second $K^2$ term is the square of the vector constraint~$-\frac1{m^4}\, {\cal C}_\mu \ell^{\mu\nu} \ell_{\nu\rho} {\cal C}^\rho$, we have the modified constraint 
\begin{equation*}
\begin{split}
0&={\cal C}-\frac1{2m^2}\, ({\cal C}.\ell_\nu)^2\\&=-\frac{3m^4}{2} \, f - \frac{m^2}{2} \, e^\nu{}_me^\mu{}_n R(f)_{\mu\nu}{}^{mn} +\frac{m^2}2 \, K_{\mu\nu\rho}K^{\nu\rho\mu}\, .
\end{split}
\end{equation*}
The first term is the familiar FP-trace and the second background curvature one vanishes for flat $f_\mu{}^m$. We will see in the next Section that it is the third, $K^2$, term which has dire consequences for the causality of the model. While it does vanish for special solutions whose contorsion obeys $K_{\mu\nu\rho}-K_{\rho\nu\mu}=0$, imposing this condition as an additional constraint would remove further   field theoretical DoF, an obviously unacceptable tradeoff.

\section{Acausality}

We study the causality of the model via its characteristics, using a  
method first introduced in a field theoretical context in~\cite{Velo,Madore}. This allows us to determine the maximum speed of propagation by studying a shock whose second derivatives are discontinuous across its wavefront. Since the model is second order in derivatives, we assume that the dynamical metric $g_{\mu\nu}$ and its first derivatives are continuous across the hypersurface spanned by the shock's wavefront by $\Sigma$. 
The inert $f_\mu{}^m$ background is of
course continuous. 
Note that we are studying causality with
respect to the dynamical metric~$g$, not the background,
this being a putative theory of the metric field. (Actually, our conclusions are equally valid with respect to the background metric.)
Then $g$, being smooth
across~$\Sigma$, defines local light-cones that allow us to decide whether the shock wavefront corresponds to superluminal propagation.

To start, we denote the leading discontinuity in the metric across~$\Sigma$ by square brackets
$$
\big[\partial_\alpha\partial_\beta g_{\mu\nu}\big]_\Sigma=\xi_\alpha \xi_\beta \gamma_{\mu\nu}\, ,
$$
where~$\xi_\mu$ is a vector normal to the characteristic and~$\gamma_{\mu\nu}$ is some nonvanishing symmetric tensor defined on the characteristic surface. 
Propagation is acausal whenever the field equations admit characteristics with timelike normal~$\xi_\mu$,~{\it i.e.},
$$
\xi^\mu g_{\mu\nu} \xi^\nu < 0\, ;
$$
it can be analyzed by studying the field equations and any combinations of field equations and their derivatives that are of degree two or less in derivatives on $g_{\mu\nu}$ and so have a well-defined discontinuity across~$\Sigma$. This, of course, amounts to studying the discontinuity of ${\cal G}_{\mu\nu}$ and the constraints ${\cal C}_\mu$ and ${\cal C}$ across~$\Sigma$. 

First, we consider the anti-symmetric part of the equation of motion~${\cal G}_{\mu\nu}$  implying $f_{\mu\nu}=f_{\nu\mu}$. For this we must compute the  discontinuity of the vierbein. Since these depend algebraically on the metric we have
$$
\big[\partial_\alpha\partial_\beta e_{\mu}{}^m\big]_\Sigma=\xi_\alpha \xi_\beta {\cal E}_{\mu}{}^m\, ,
$$
where ${\cal E}_{\mu}{}^m$ is some tensor defined on the characteristic surface. Computing the discontinuity of the relation $e_{\mu}{}^m \eta_{mn} e_\nu{}^n=g_{\mu\nu}$ gives
$
\xi_{\alpha}\xi_\beta\big({\cal E}_{\mu\nu} + {\cal E}_{\nu\mu} \big)= \xi_{\alpha}\xi_\beta \gamma_{\mu\nu}
$. 
At this point, we proceed by contradiction by taking $\xi_\mu$ timelike. Without loss of generality, we may therefore set
$$
\xi^\mu g_{\mu\nu} \xi^\nu=-1\, ,
$$
and thus learn that
$$
{\cal E}_{\mu\nu} + {\cal E}_{\nu\mu} = \gamma_{\mu\nu}\, .
$$
A similar computation based on the symmetry of $f_{\mu\nu}$ gives
\begin{equation}\label{fE}
f_\mu{}^\rho {\cal E}_{\nu\rho}=f_{\nu}{}^\rho {\cal E}_{\mu \rho}\, .
\end{equation}
Next we compute the leading discontinuity in the field equation ${\cal G}_{\mu\nu}$ and in turn its trace ${\cal G}$. Since this amounts to studying the second derivative terms in these equations, the result coincides with that of the FP theory computed long ago in~\cite{Shamaly, DW} (save that indices are raised and lowered with the metric $g_{\mu\nu}$):
\begin{eqnarray}
\xi^2 \gamma_{\mu\nu} -\xi_\mu \xi.\gamma_\nu-\xi_\nu \xi.\gamma_\mu
+\xi_\mu\xi_\nu\, \gamma &=&0\, ,\label{deoms}
\\[2mm]
\xi^2 \gamma -\xi.\xi.\gamma &=&0\, .\nonumber\end{eqnarray}
It is clearly useful to decompose our
variables with respect to the 
 (unit) timelike vector~$\xi_\mu$. In particular, for a  vector, symmetric tensor and 
antisymmetric tensor we have, respectively, 
$$
\begin{array}{cclr}
V_\mu&\!\!\!:=\!\!&V^\perp_\mu -\xi_\mu \xi.V\, ,\\[1mm]
\!\!S_{\mu\nu}&\!\!\!:=\!\!&
S^\perp_{\mu\nu}\!-\xi^\noperp_\mu S_\nu^\perp -\xi^\noperp_\nu S_\mu^\perp \!+ \xi_\mu\xi_\nu\,  \xi.\xi.S\, ,&\hspace{-1mm}\big(S_\mu:=\xi.S_\mu\big) ,\\[2mm]
A_{\mu\nu}&\!\!\!:=\!\!&A^\perp_{\mu\nu}+\xi^\noperp_\mu A_\nu^\perp -\xi^\noperp_\nu A_\mu^\perp\, ,&\hspace{-13mm}\big(A^\perp_\mu:=A_{\mu\nu} \xi^\nu\big) .\\
\end{array}
$$
In this language, Eq.~(\ref{deoms}) implies that $\gamma^\perp_{\mu\nu}=0$ so 
\begin{equation}\label{gamma}
\gamma_{\mu\nu}=-\xi^\noperp_\mu \gamma_\nu^\perp -\xi^\noperp_\nu \gamma_\mu^\perp + \xi_\mu\xi_\nu\,  \xi.\xi.\gamma\, .
\end{equation}
The next task is to compute the discontinuity in the vector constraint:
\begin{eqnarray*}
\big[\xi^\alpha \partial_\alpha {\cal C}_\mu\big]_\Sigma&=&\!\phantom{-}m^2\,  \xi^\alpha\big[ \partial_\alpha\omega(e)_\rho{}^{\rho\sigma}\big]_\Sigma\, f_{\mu\sigma}
\\[2mm]&=&\!-m^2 \big({\cal E}_\nu{}^\nu  \xi^\sigma- \, {\cal E}^{\sigma\nu} \xi_\nu\big)\, f_{\mu\sigma}\, .
\end{eqnarray*}
Since $f_{\mu\nu}$ is assumed to be invertible, by decomposing $$2{\cal E}_{\mu\nu}=\gamma_{\mu\nu}+a_{\mu\nu}\, ,$$
into its symmetric and antisymmetric parts, we find
\begin{equation}\label{a}
0=\gamma^\perp_\mu + a^\perp_\mu\, .
\end{equation}
Together, Equations~(\ref{gamma}) and~(\ref{a}) give $2{\cal E}_{\mu\nu}=a^\perp_{\mu\nu}-2\xi_\mu^\noperp \gamma^\perp_\nu + \xi_\mu\xi_\nu \, \xi.\xi.\gamma$
so that Eq.~(\ref{fE}) becomes
\begin{equation}\label{fdep}
0=f^\perp_\mu{}^\rho a^\perp_{\nu\rho}
+\xi_\mu\big(2f^\perp_\nu{}^\rho\gamma^\perp_\rho-f_\rho^\perp a^\perp_\nu{}^\rho-\xi.\xi.\gamma \, f_\nu^\perp\big)
-\big(\mu\leftrightarrow\nu\big)\, .
\end{equation}
The terms perpendicular and parallel to $\xi_\mu$ must vanish separately, so 
\begin{equation}\label{feqs}
f^\perp_\mu{}^\rho a^\perp_{\nu\rho}-f^\perp_\nu{}^\rho a^\perp_{\mu\rho}=0=2f^\perp_\nu{}^\rho\gamma^\perp_\rho-f_\rho^\perp a^\perp_\nu{}^\rho-\xi.\xi.\gamma \, f_\nu^\perp\, .
\end{equation}
The first set of these equations {\it generically} gives three independent linear conditions on as many unknowns ($a_{\mu\nu}^\perp$) so enforces $a^\perp_{\mu\nu}=0$.
The second set then gives three conditions on the four remaining nonvanishing unknowns, $\gamma^\perp_\mu$ and $\xi.\xi.\gamma$. Thus, {\it generically}  three linear combinations
of these vanish, leaving one nonzero linear combination. If this were to vanish, we would have established the absence of shock wavefronts $\Sigma$ with timelike normal~$\xi_\mu$.
(Of course, one still would have to verify the absence of special cases for the two italicized appearances of ``generically'' in the preceding argument; those are irrelevant in the face of the generic acausality we are about to exhibit.)

At this stage, then, the model is left requiring one more condition on ${\cal E}_\mu{}^m$ for its causal consistency. That condition can only derive from the remaining scalar constraint~${\cal C}$, whose 
discontinuity across~$\Sigma$ we  compute next. To begin with, to better exhibit the problem we are about to find, let us make the assumption that the background is flat and 
that the contorsion vanishes so that the remaining constraint implies $f=0$ whose discontinuity across~$\Sigma$ implies $f^{\mu\nu}{\cal E}_{\mu\nu}=0$. This provides the remaining 
independent linear relation between $\xi.\xi.\gamma$ and $\gamma_\mu^\perp$ required to establish that ${\cal E}_{\mu}{}^m=0$ and in turn the absence of superluminal shocks--{\it so long as the contorsion vanishes}.

However, the contorsion does not vanish as a consequence of the field equations (in fact, as discussed above this would imply too many conditions on the field theoretic DoF).
Thus a proper computation of the discontinuity of ${\cal C}$ reads
\begin{eqnarray*}
\big[\xi^\alpha\partial_\alpha\big({\cal C}-\frac1{2m^2}\, ({\cal C}.\ell_\nu)^2\big)\big]_\Sigma\!&=\!&\phantom{-}\frac{m^2}2\, \xi^\alpha \big[\partial_\alpha\big(K_{\mu\nu\rho}K^{\nu\rho\mu}\big)\big]_\Sigma\\[1mm]
&=\!&-\frac{m^2}{2}\, \xi_\nu K^{\mu\nu\rho}{\cal E}_{\rho\mu}\\[1mm]
&=\!&\phantom{-}\frac{m^2}{4}\, \xi_\nu^\noperp K^{\mu\nu\rho} a^\perp_{\mu\rho}\, .
\end{eqnarray*}
Thus, instead of a relation involving $\xi.\xi.\gamma$ and $\gamma_\mu^\perp$, we find the seemingly additional, but in fact redundant, requirement $\xi_\nu^\noperp K^{\mu\nu\rho} a^\perp_{\mu\rho}=0$ on~$a^\perp_{\mu\nu}$. Therefore, since some linear combination of $\xi.\xi.\gamma$ and~$\gamma_\mu^\perp$ does not vanish,  timelike shock normals are allowed. This establishes the promised presence of acausal characteristics for any choice of background.

\section{Discussion}

We have just shown that one otherwise ghost-free, acceptable finite
range gravity model is excluded. How far does this no-go result
extend to all three possible such combinations, quite apart from other
previously mentioned obstacles to these models? 
Very recently, causality for models 
with mass terms quadratic in~$f$ has been ruled out~\cite{DSW} using methods similar to the present ones.
This leaves only the third candidate  mass term, cubic in~$f$:
Any model of the form $G_{\mu\nu}(g) = T_{\mu\nu}(f,e)$ with algebraic~$T$  universally yields 
Eq.~(\ref{gamma}) for the shock; the structure of the fifth constraint is at the root of the acausality~\footnote{Given the scalar constraint's nefarious role, one might try
to turn it into a harmless Bianchi identity
by taking a de  Sitter background  and a partially massless (PM) limit where the scalar helicity does
not propagate~\cite{PM}. However, very recently it has been shown that no PM limit of ghost-free massive $f$-$g$ theories exists~\cite{DSW} (a result just reconfirmed in~\cite{dRPM}).}. Its covariant version for the third mass
term is as yet unknown, but
 if it takes the generic form $f^3 + f^2 K^2$ where $K$ is the contorsion, the argument of~\cite{DSW} already establishes its acausality.
Even if it does not, there is a potentially  new source of discontinuity, closer to that of the charged massive spin~3/2 and~2 systems~\cite{Velo,Shamaly,DW,Madore}. 
Namely, zeros in the characteristic matrix can allow superluminal characteristics, just as critical values of the background {\it  E/M} field permit superluminal signal propagation in the charged $s=(3/2,2)$ models. 
In fact, for those models, acausality can be traced to nonpositivity of equal time commutators, a fatal physical flaw~\cite{JS}. We conclude therefore that the acausality we have exhibited is
an unavoidable pathology of $f$-$g$ massive gravity barring some miracle of the cubic model or 
some (hitherto unknown) underlying ``rescue'' modification~\footnote{One example, in a different context, is the use of  string theoretically-inspired  nonminimal couplings for charged higher spins~\cite{Porrati}.}  that also yields a smooth massless limit~\footnote{The massless vDVZ discontinuity can actually be averted by 
introducing a cosmological constant and setting the mass to zero before limiting to flat space~\cite{Lambda}. 
Also, it was suggested long ago   that a similar  mechanism applies to the interchange of massless and free limits in a putative nonlinear massive theory~\cite{Vainshtein}.}. 
Indeed, the fact that neither GR nor Yang--Mills have massive ``neighbors'' is a self-sharpening Occam's razor that further ornaments these fundamental pillars!

\begin{acknowledgments}
We  thank M. Sandora for a very useful discussion; A.W. acknowledges the hospitality of the Lauritsen Lab, Caltech. S.D. was supported in part by NSF Grant No. PHY- 1064302 and DOE Grant No. DE-FG02-164 92ER40701. 
\end{acknowledgments}


\end{document}